\documentclass[aps,pre,twocolumn,groupedaddress]{revtex4}

\usepackage{graphicx}
\usepackage{dcolumn}
\usepackage{bm}
\usepackage{subfigure}

\begin{document}


\title{Scaling Behavior in the 3D Random Field $XY$ Model}


\author{Ronald Fisch}
\email[]{ronf124@gmail.com}
\affiliation{382 Willowbrook Dr.\\
North Brunswick, NJ 08902}


\date{\today}

\begin{abstract}
We have performed studies of the 3D random field $XY$ model on
$L \times L \times L$ simple cubic lattices with periodic boundary conditions,
with a random field strength of $h_r$ = 1.875, for $L = 64$ and $L = 96$,
using a parallelized Monte Carlo algorithm.  We present results for the
angle-averaged magnetic structure factor, $S ( k )$ at $T = 1.00$, which
appears to be the temperature at which small jumps in the magnetization per
spin and the energy per spin occur.  The results indicate the existence of an
approximately logarithmic divergence of $S ( k )$ as $k \to 0$.  This suggests
that the lower critical dimension for long range order in this model is three.

\end{abstract}


\maketitle

\section{Introduction}

The behavior of the three-dimensional (3D) random-field $XY$ model
(RFXYM) at low temperatures and weak to moderate random field
strengths continues to be controversial.  A detailed calculation by
Larkin\cite{Lar70} showed that, in the limit that the number of
spin components, $n$, becomes infinite, the ferromagnetic phase
becomes unstable when the spatial dimension of the lattice is less
than or equal to four, $d \le 4$.  Dimensional reduction
arguments\cite{IM75,AIM76} appeared to show that the long-range order
is unstable for $d \le 4$ for any finite $n \ge 2$.  However, there
are several reasons for questioning whether dimensional reduction
can be trusted for $XY$, {\it i.e.} $n = 2$, spins.

Some time ago, Monte Carlo calculations\cite{GH96,Fis97} showed
that there was a line in the temperature vs. random-field plane
of the phase diagram of the three-dimensional (3D) random-field
$XY$ model (RFXYM), at which the magnetic structure factor becomes
large as the wave-number $k$ becomes small.  Additional calculations\cite{Fis07}
indicated that there appeared to be small jumps in the magnetization
and the energy of $L = 64$ lattices at a random field strength of
$h_r = 2.0$, at a temperature somewhat below $T = 1.0$.  Further
calculations\cite{Fis10} showing similar behavior for other values
of the random field strength were also performed. If such behavior
persists for larger values of $L$, this would demonstrate that there
is a ferromagnetic phase at weak to moderate random fields and low
temperatures for this model.

Since there have been substantial improvements in computing hardware
and software over the last ten years, the author felt it worthwhile
to conduct a new Monte Carlo study of this model using parallel
processing.  The results of that study for $L = 64$ and $L = 96$ will
be presented here.  The extension of the methods used here to $L = 128$
lattices is currently in progress.

\section{The Model}

For fixed-length classical spins the Hamiltonian of the RFXYM is
\begin{equation}
  H ~=~ - J \sum_{\langle ij \rangle} \cos ( \phi_{i} - \phi_{j} )
  ~-~ h_r \sum_{i} \cos ( \phi_{i} - \theta_{i} )  \, .
\end{equation}
Each $\phi_{i}$ is a dynamical variable which takes on values
between 0 and $2 \pi$. The $\langle ij \rangle$ indicates here a sum
over nearest neighbors on a simple cubic lattice of size $L \times L
\times L$. We choose each $\theta_{i}$ to be an independent
identically distributed quenched random variable, with the
probability distribution
\begin{equation}
  P ( \theta_i ) ~=~ 1 / 2 \pi   \,
\end{equation}
for $\theta_i$ between 0 and $2 \pi$.  We set the exchange constant to
$J = 1$, with no loss of generality.  This Hamiltonian is closely
related to models of vortex lattices and charge density waves.\cite{GH96,Fis97}

Larkin\cite{Lar70} studied a model for a vortex lattice in a
superconductor.  His model replaces the spin-exchange term of the
Hamiltonian with a harmonic potential, so that each $\phi_{i}$ is no
longer restricted to lie in a compact interval.  He argued that for
any non-zero value of $h_r$ this model has no ferromagnetic phase on
a lattice whose dimension $d$ is less than or equal to four.  The Larkin
approximation is equivalent to a model for which the number of spin
components, $n$, is sent to infinity.  A more intuitive derivation
of this result was given by Imry and Ma,\cite{IM75} who assumed that
the increase in the energy of an $L^d$ lattice when the order parameter
is twisted at a boundary scales as $L^{d - 2}$ for all $n > 1$, just as
it would for $h_r = 0$.  Using this assumption, they argued that when
$d \le 4$ there is a length $\lambda$, now called the Imry-Ma length,
at which the energy which can be gained by aligning a spin domain with
its local random field exceeds the energy cost of forming a domain wall.
From this they claimed that the magnetization would decay to zero when
the system size, $L$, exceeds $\lambda$.

Within a perturbative $\epsilon$-expansion one finds the phenomenon
of ``dimensional reduction". The critical exponents of any
$d$-dimensional $O(n)$ random-field model appear to be identical to
those of an ordinary $O(n)$ model of dimension $d - 2$. For the
$n = 1$ (RFIM) case, this was soon shown rigorously to be incorrect for
$d < 4$.\cite{Imb84,BK87}  More recently, extensive numerical results
for the Ising case have been obtained for $d = 4$ and
$d=5$.\cite{FMPS17a,FMPS17b}  They determined that dimensional reduction is
ruled out numerically in the Ising case for $d = 4$, but not for $d = 5$.\cite{FMPS17c}

Because translation invariance is broken for any non-zero $h_r$, it
seems quite implausible to the current author that the twist energy for
Eqn.~(1) scales as $L^{d - 2}$, even though this is correct to all orders
in perturbation theory.  An alternative derivation by Aizenman and
Wehr,\cite{AW89,AW90} which claims to be mathematically rigorous,
also makes an assumption equivalent to translation invariance.
Although the average over the probability distribution of random
fields restores translation invariance, one must take the infinite
volume limit first.  It is not correct to interchange the infinite
volume limit with the average over random fields.  This problem of
the interchange of limits is equivalent to the existence of replica
symmetry breaking.  The existence of replica symmetry breaking in
random field models was first shown by Mezard and Young,\cite{MY92}
about two years after the work of Aizenman and Wehr.  Mezard and
Young emphasized the Ising case, and the fact that this applies for
all finite $n$ seems to have been overlooked by many people for a
number of years.  A functional renormalization group calculation
going to two-loop order was performed by Tissier and Tarjus,\cite{TT06}
and independently by Le Doussal and Wiese.\cite{LW06}  They found
that there was a stable critical fixed point of the renormalization
group for some range of $d$ below four dimensions in the $n = 2$
random field case.  However, it is not clear from their calculation
what the nature of the low-temperature phase is, or whether this
fixed point is stable down to $d = 3$.  Tarjus and Tissier\cite{TT08}
later presented an improved version of this calculation, which
explains more explicitly why dimensional reduction fails for the
$n = 2$ case when $d \leq 4$.

\section{Structure factor}

The magnetic structure factor, $S (\vec{\bf k}) = \langle
| \vec{\bf M}(\vec{\bf k}) |^2 \rangle $, for $XY$ spins is
\begin{equation}
  S (\vec{\bf k}) ~=~  L^{-3} \sum_{ i,j } \cos ( \vec{\bf k} \cdot
  \vec{\bf r}_{ij}) \langle \cos ( \phi_{i} - \phi_{j}) \rangle  \,   ,
\end{equation}
where $\vec{\bf r}_{ij}$ is the vector on the lattice which starts
at site $i$ and ends at site $j$, and here the angle brackets denote
a thermal average.  For a random field model, unlike a random bond
model, the longitudinal part of the magnetic susceptibility, $\chi$,
which is given by
\begin{equation}
  T \chi (\vec{\bf k}) ~=~ 1 - M^2 ~+~ L^{-3} \sum_{ i \ne j } \cos (
  \vec{\bf k}  \cdot \vec{\bf r}_{ij}) (\langle \cos ( \phi_{i} - \phi_{j}
  ) \rangle ~-~ Q_{ij} )  \,   ,
\end{equation}
is not the same as $S$ even above $T_c$.  For $XY$ spins,
\begin{equation}
  Q_{ij} ~=~ \langle \cos ( \phi_{i} ) \rangle \langle \cos (
  \phi_{j} ) \rangle ~+~ \langle \sin ( \phi_{i} ) \rangle \langle
  \sin ( \phi_{j} ) \rangle  \,  ,
\end{equation}
and
\begin{equation}
  M^2 ~=~ L^{-3} \sum_{i} Q_{ii}
      ~=~ L^{-3} \sum_{i} \langle \cos ( \phi_{i} ) \rangle^2 ~+~
       \langle \sin ( \phi_{i} ) \rangle^2  \,  .
\end{equation}
When there is a ferromagnetic phase transition, $S ( \vec{\bf k} = 0 )$
has a stronger divergence than $\chi ( \vec{\bf k} = 0 )$.

The scalar quantity $\langle M^2 \rangle$, when averaged over a set
of random samples of the random fields, is a well-defined function
of the lattice size $L$ for finite lattices. With high probability,
it will approach its large $L$ limit smoothly as $L$ increases.
The vector $\vec{\bf M}$, on the other hand, is not really a
well-behaved function of $L$ for an $XY$ model in a random field.
Knowing the local direction in which $\vec{\bf M}$ is pointing,
averaged over some small part of the lattice, may not give us a
strong constraint on what $\langle \vec{\bf M} \rangle$ for the
entire lattice will be.  When we look at the behavior for
all $k$, instead of merely looking at $k = 0$, we get a much better
idea of what is really happening.

\section{Numerical results for $S ( k )$}

In this work, we will present results for the average over angles of
$S (\vec{\bf k})$, which we write as $S (k)$.  The data were
obtained from $L \times L \times L$ simple cubic lattices with $L =
64$ and $L = 96$ using periodic boundary conditions.  The calculations
were done using a clock model which has 8 equally spaced dynamical
states at each site.  In addition, there is a static random phase at
each site which was chosen to be $0, \pi/12$ or $\pi/6$ with equal
probability.  It has been known for some time that a model of this type,
without the random-field term, is in the universality class of the pure
$XY$ model under most conditions, even if the number of dynamical states
of each spin is only 3.\cite{Fis92}  Under conditions of very low
temperature, this model may undergo an incommensurate-to-commensurate
type of charge-density wave phase transition.  Thus it is expected that,
when we include the random-field term, the model will behave essentially
as a random-field $XY$ model, as long as we do not attempt to work at
very low temperature.\cite{Fis97}

The strength of the random field for which data were obtained is
$h_r$ = 1.875.  This value was chosen in order to make the value of
$T_c$ close to 1.00.  The direction of the random field at site $i$,
$\theta_i$, was chosen randomly from the set of the 24th roots of unity,
independently at each site.  Since $\theta_i$ has 24 possible values,
our past experience with models of this type indicates that there is no
possibility that the discretization will affect the behavior near
$T = 1.00$ in an observable way.

The computer program uses three independent pseudorandom number generators:
one for choosing the values of the $\theta_i$, one for setting the
static random phases, and a third one for the Monte Carlo spin flips,
which are performed by a single-spin-flip heat-bath algorithm.  The
pseudorandom number generator used for the Monte Carlo spin flips
was the library function $random\_number$, supplied by the Intel
Fortran compiler, which is suitable for parallel computation.  The
spin-flip subroutine was parallelized using OpenMP, by taking
advantage of the fact that the simple cubic lattice is two colorable.
It was run on Intel multicore processors of the Bridges Regular
Memory machine at the Pittsburgh Supercomputer Center.  The code was
checked by setting $h_r = 0$, and seeing that the known behavior
of the pure ferromagnetic 3D XY model was reproduced correctly.

24 different realizations of the random fields $\theta_i$ were
studied for $L = 64$, and another 24 samples were studied for $L = 96$.
Each lattice was started off in a random spin state at $T = 2.25$,
above the $T_c$ for the pure model, and cooled slowly to $T = 1.00$.
At $T = 1.00$, the sample was relaxed until an apparent equilibrium
was reached over an appropriate time scale.  For $L = 64$ this time
scale was 163,840 Monte Carlo steps per spin (MCS), and for $L = 96$
this was increased to at least 655,360 MCS.  Some $L = 96$ samples
required relaxation for up to three times longer.

After each sample was relaxed at $T = 1.00$, a sequence of 8 equilibrated
spin states obtained at intervals of 20,480 MCS for $L = 64$ or 40,960 MCS
for $L = 96$ was Fourier transformed to calculate $S (\vec{bf k})$, and
then averaged over the sequence of 8 spin states. The data were then
binned according to the value of $k^2$, to give the angle-averaged $S ( k )$.
Finally, an average over the 24 samples was performed for each $L$.  The
average magnetization per spin at $T = 1.00$ of these slowly cooled samples
was $0.139 \pm 0.011$ for $L = 64$, and $0.080 \pm 0.011$ for $L = 96$.

Data were also obtained for the same sets of samples using ordered
initial states and warming to $T = 1.00$.  At least two, and sometimes
more initial ordered states were used for each sample.  The initial
magnetization directions used were chosen to be close to the direction
of the magnetization of the slowly cooled sample with the same set of
random fields.  This type of initial state was chosen because it was
found in the earlier work\cite{Fis07} that this is the way to find the
lowest energy minima in the phase space.  The data from the initial
condition which gave the lowest average energy for a given sample was
then selected for further analysis and comparison with the slowly cooled
state data from that sample.  The relaxation procedure at $T = 1.00$
for the warmed states was the same one used for the cooled states, and
the calculation of $S (k)$ proceeded in the same way.  The average
magnetization per spin of these selected warmed states was $0.239 \pm 0.010$
for $L = 64$, and $0.157 \pm 0.011$ for $L = 96$.

\begin{figure}
\includegraphics[width=3.4in]{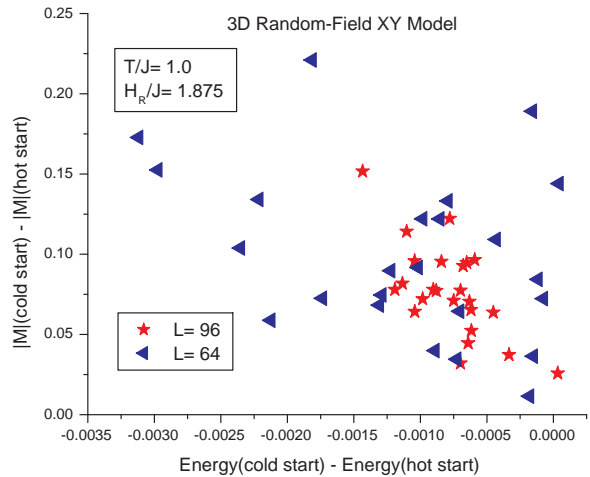}
\caption{\label{Fig.1}(color online) Jump in the magnetization vs.
jump in the energy for $L = 64$ and $L=96$ simple cubic lattices
with $h_r = 1.875$ at $T = 1.00$.  States with hot start and ordered
start initial conditions are compared for each sample.}
\end{figure}

In order to compare the $L = 64$ data with the $L = 96$ data, the energy
per spin difference and the magnetization per spin difference between the
cooled state and the warmed state at $T = 1.00$ were computed for each
sample.  The results are shown in a scatter plot in Fig.~1. We see that
the distributions do not show any significant correlation between the
energy difference and the magnetization difference for the $L = 64$ samples.
For the $L = 96$ samples there is a weak tendency for the size of the
jump in the magnetization to be correlated with the size of the jump in
the energy. The distribution is rather broad for $L = 64$, and significantly
narrower for $L = 96$.

The center of the $L = 64$ distribution is at $\delta |M| = 0.100 \pm 0.011$
and $\delta E = -0.00113 \pm 0.00019$, while the center of the $L = 96$
distribution is at $\delta |M| = 0.077 \pm 0.006$ and $\delta E = -0.00078
\pm 0.00006$.  The conjecture that $\delta |M|$ and $\delta E$ will scale to
zero\cite{Fis07} as $L \to \infty$ is consistent with these data.  There is
some indication that $\delta |M|$ scales more slowly than $\delta E$, but more
data are needed before any quantitative estimate of the rates of convergence
of these parameters can be made.  The author is planning to make such
estimates when data for $L = 128$ samples become available.

The average specific heat of the $L = 64$ samples at $T = 1.00$ is
$0.7797 \pm 0.0019$ for the cooled samples, and $0.7815 \pm 0.0028$
for the heated samples.  The corresponding numbers for $L = 96$ are
$0.7802 \pm 0.0028$ and $0.7854 \pm 0.0026$.  The fact that the specific
heat of the cooled samples is lower than the specific heat of the
somewhat more magnetized heated samples is expected.  The fact that
the difference between them is very small means that there is not much
energy associated with the disappearance of the magnetic long-range
order.  The fact that the jump in the specific heat seems to be slightly
larger for $L = 96$ than for $L = 64$ is normal for a weakly first-order
phase transition.  The fact that the jump is so small also means that
we are not looking at a normal second order phase transition.

The uncertainty in our estimate of the $T_c$, the temperature of the
phase transition, is about an order of magnitude less than the extrapolated
shift in temperature which would be needed to make the jump in energy
between the heated samples and the cooled samples disappear.  However, the
free-energy minimum of a cold-start sample actually becomes clearly
unstable at a temperature a few percent higher than $T = 1.00$.

Now we turn to the data for the structure factor.  The average $S ( k )$
for the 24 $L = 96$ samples at $T = 1.00$ is shown in Fig.~2.  $S ( k )$
is computed separately for the heated sample data and the cooled sample
data, but it is difficult to see any difference between them.  These
data are very similar to the earlier data\cite{Fis07} for $h_r =2$ at
$T = 0.875$.  The change in the slope of the data points now occurs near
$k = 0.11$ instead of $k = 0.14$, but this is about what is expected from
using the somewhat lower value of $h_r$.  From this log-log plot, it is
not clear how to extrapolate the data to small $k$.

\begin{figure}
\includegraphics[width=3.4in]{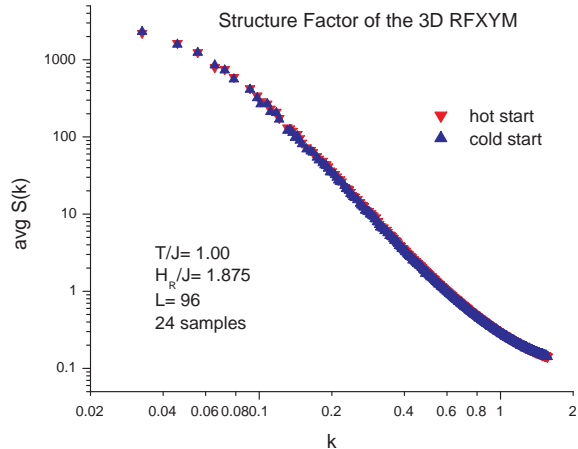}
\caption{\label{Fig.2}(color online) Angle-averaged structure factor
vs. $k$ for $96 \times 96 \times 96$ lattices with $h_r = 1.875$ at
temperature T = 1.00. Both the $x$-axis and the $y$-axis are scaled
logarithmically.  One $\sigma$ statistical errors are approximately
the size of the plotting symbols.}
\end{figure}

\begin{figure}
\includegraphics[width=3.4in]{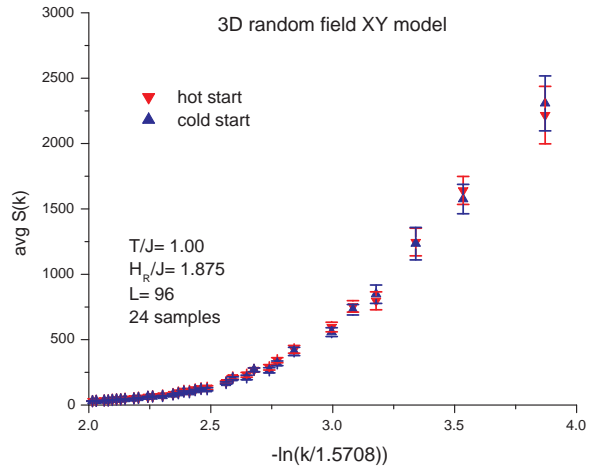}
\caption{\label{Fig.3}(color online) Angle-averaged structure factor
vs. (-~ln(k/1.5708)) for $96 \times 96 \times 96$ lattices with
$h_r = 1.875$ at temperature T = 1.00.  One $\sigma$ statistical errors
are shown.}
\end{figure}

To clarify the behavior at small $k$, we replot the same $L = 96$
data for the structure factor on a linear scale in Fig.~3.  The
scaling for the $x$-axis is chosen so that the edge of the Brillouin
zone would be at $x = 0$, but only the small-$k$ part of the data are
shown on the graph.  From Fig.~3 it is clear that we have no evidence
for a finite correlation length at $L = 96$.  However, these data also
appear to rule out the possibility that $S ( k )$ behaves like
$k^{4 - \bar{\eta}}$ with $1 < \bar{\eta} \le 2$ as $k \to 0$, which
would be required for hyperscaling to hold.

A divergence of $S ( k )$ as $k \to 0$ like $\ln (k)$, or some power
of $\ln (k)$, is a strong indicator that the lower critical dimension
of the RFXYM is exactly equal to three.  The author is not aware of
another example of this type of behavior in a model with quenched
random disorder, and much remains to be learned.  It would be very
exciting if similar behavior was observed by doing experiments on
physical systems which are believed to be in the universality class
of this model.

\section{Discussion}

It is straightforward to calculate the interaction energy of the
spins with the random field.  We merely need to calculate the value
of the second sum in the Hamiltonian as a function of the temperature.
When this is done at $h_r = 1.875$, it turns out that the value of
the random-field energy has a maximum at about $T = 1.75$  Below that
temperature, the ferromagnetic bonds become increasingly successful
in pulling the directions of the local spins away from the directions
of their local random fields.  Of course, there is nothing magic about
$T = 1.75$.  The temperature at which the maximum value in the
random-field energy will occur will be a function of the value of $h_r$.
This effect is not accounted for in the Imry-Ma argument.

Finding that $S (k)$ diverges at low temperatures in the RFXYM as
$k \to 0$ is not surprising.  This behavior follows from the results of
Aharony\cite{Aha78} for models which have a probability distribution
for the random fields which is not isotropic.  According to Aharony's
calculation, if this distribution is even slightly anisotropic, then
we should see a crossover to RFIM behavior.  We know\cite{Imb84,BK87}
that in $d = 3$ the RFIM is ferromagnetic at low temperature if the
random fields are not very strong.  The instability to even a small
anisotropy in the random field distribution should induce a diverging
response in $S (k)$ as $k \to 0$ for the RFXYM in $d = 3$.  A similar
effect in a related, but somewhat different, model was found by Minchau
and Pelcovits.\cite{MP85}.

There has been no attempt in this work to equilibrate samples at
temperatures below the apparent critical temperature.  Therefore we
have no data which directly address the question of whether the RFXYM
shows true ferromagnetism in $d = 3$.  If we assume that the average
$|M|$ of finite samples is subextensive, i.e. the net magnetic moment
grows more slowly than $L^3$ as $L \to \infty$, then it would follow
from the above argument that there should not be any divergence of
$S (k)$ for $k \to 0$ in $d = 3$ in the cases $n \ge 3$.  If this were
the case, then the behavior of random field $O (n)$ models in $d = 3$
would be remarkably parallel to the case of the ordinary $O (n)$
ferromagnets in $d = 2$.  One might hope to find a relatively simple
reason for such an effect.

About five years ago, numerical studies of the RFXYM were performed
by Garanin. Chudnovsky and Proctor.\cite{GCP13}  These authors were
interested in studying lattices of very large $L$.  Such lattices
were much too large for the simulations to be able to reach a thermal
equilibrium, and they did not use any Boltzmann factors in their
dynamics.  Thus the results are some kind of simulated annealing, and
it is not clear what the meaning of their end states is.  The work
being reported here always used Boltzmann factors to relax the state
of the lattice.  It is not possible to make any quantitative comparison,
because they only study low energy states, and give no results for the
behavior at the phase transition.  In further work,\cite{PGC14} these
authors extend their methods to models with other numbers of spin components.
They claim that the 3D $n = 3$ spin model in a random field of $h_r = 1.5$
also has a stable ferromagnetic phase at low temperature, but do not
give an estimate of $T_c$.  They also claim that for 2D, the RFXYM
has a ferromagnetic state for $h_r = 0.5$, which is surely incorrect.
Therefore the reliability of their methods is highly questionable.  The
functional renormalization group calculations\cite{TT06,LW06,TT08} do not
give any support for the existence of a ferromagnetic phase for the $n = 3$
random field case for $d \leq 4$.

There is another model which is more similar to the random-field $XY$
model than the random-field Ising model is.  That model is the 3-state
Potts model in a random field (RFPM).  In the absence of the random-field
term, a 3D 3-state Potts model has a first-order phase transition,
with a substantial latent heat at $T_c$.  In 1989, two groups presented
independent arguments showing that models like this should no longer
have a latent heat when the random-field term is added to the Hamiltonian.
Aizenman and Wehr\cite{AW89} proved that the latent heat must vanish
in the limit $L \to \infty$.  Hui and Berker\cite{HB89} argued that the
vanishing of the latent heat implied that a critical fixed point should
exist.  This author does not see, however, why such a fixed point, with
its associated divergent correlation length, should generally exist in a
model which has no translation symmetry, except in those cases where the
randomness is an irrelevant operator.\cite{Har74}  It is certainly true
that there are some cases where such fixed points have been found using
$\epsilon$-expansion calculations.  Subextensive singularities in the
specific heat and the magnetization are completely consistent with the
Aizenman-Wehr Theorem.

\section{Summary}

In this work we have performed Monte Carlo studies of the 3D RFXYM on
$L = 64$ and $L = 96$ simple cubic lattices, with a random field strength of
$h_r = 1.875$.  We compare the properties of slowly cooled states and slowly
heated states at $T = 1.00$, which is our estimate of the temperature at which
there appears to be a phase transition.  We display results for the change in
energy and the change in magnetization at this temperature, as a function of
the lattice size.  At the phase transition we measure small jumps in the
magnetization per spin and the energy per spin.  However, it is likely that
these jumps are subextensive, meaning that they probably scale to zero as
$L \to \infty$.  We also compute results for the structure factor, $S ( k )$,
under these conditions.  For $L = 96$ the structure factor appears to be have
an approximately logarithmic divergence in the small $k$ limit.  These
characteristics are consistent with the idea that the lower critical dimension
of this model is exactly three.

\begin{acknowledgments}
The author thanks N. Sourlas for a helpful conversation about the recent work
on the random field Ising model. This work used the Extreme Science and Engineering
Discovery Environment (XSEDE) Bridges Regular Memory at the Pittsburgh Supercomputer
Center through allocations DMR170067 and DMR180003.  The author thanks the staff of
the PSC for their help.

\end{acknowledgments}


\begin{thebibliography}{49}

\bibitem{Lar70}
A. I. Larkin, Zh. Eksp. Teor. Fiz. {\bf 58}, 1466 (1970) [Sov. Phys.
JETP {\bf 31}, 784 (1970)].
\bibitem{IM75}
Y. Imry and S.-K. Ma, Phys. Rev. Lett. {\bf 35}, 1399 (1975).
\bibitem{AIM76}
A. Aharony, Y. Imry and S.-K. Ma, Phys. Rev. Lett. {\bf 36}, 1364
(1976).
\bibitem{GH96}
M. J. P. Gingras and D. A. Huse, Phys. Rev. B {\bf 53}, 15193
(1996).
\bibitem{Fis97}
R. Fisch, Phys. Rev. B {\bf 55}, 8211 (1997).
\bibitem{Fis07}
R. Fisch, Phys. Rev. B {\bf 76}, 214435 (2007).
\bibitem{Fis10}
R. Fisch, arXiv:1001.3397.
\bibitem{Imb84}
J. Z. Imbrie, Phys. Rev. Lett. {\bf 53}, 1747 (1984).
\bibitem{BK87}
J. Bricmont and A. Kupiainen, Phys. Rev. Lett. {\bf 59}, 1829
(1987).
\bibitem{FMPS17a}
N. G. Fytas, V. Martin-Mayor, M. Picco and N. Sourlas, J. Stat.Mech.
033302 (2017).
\bibitem{FMPS17b}
N. G. Fytas, V. Martin-Mayor, M. Picco and N. Sourlas, Phys. Rev. E
{\bf 95}, 042117 (2017).
\bibitem{FMPS17c}
N. G. Fytas, V. Martin-Mayor, M. Picco and N. Sourlas, arXiv:1711.09597v3.
\bibitem{AW89}
M. Aizenman and J. Wehr, Phys. Rev. Lett. {\bf 62}, 2503 (1989);
erratum: Phys. Rev. Lett. {\bf 64}, 1311 (1990).
\bibitem{AW90}
M. Aizenman and J. Wehr, Commun. Math. Phys. {\bf 130}, 489 (1990).
\bibitem{MY92}
M. Mezard and A. P. Young, Europhys. Lett. {\bf 18}, 653 (1992).
\bibitem{TT06}
M. Tissier and G. Tarjus, Phys. Rev. Lett. {\bf 96}, 087202 (2006);
Phys. Rev B {\bf 74}, 214419 (2006).
\bibitem{LW06}
P. Le Doussal and K. J. Wiese, Phys. Rev. Lett. {\bf 96}, 197202 (2006).
\bibitem{TT08}
M. Tissier and G. Tarjus, Phys. Rev B {\bf 78}, 024204 (2008).
\bibitem{Fis92}
R. Fisch, Phys. Rev. B {\bf 46}, 242 (1992).
\bibitem{Aha78}
A. Aharony, Phys. Rev. B {\bf 18}, 3328 (1978).
\bibitem{MP85}
B. J. Minchau and R. A. Pelcovits, Phys. Rev. B {\bf 32}, 3081 (1985).
\bibitem{GCP13}
D. A. Garanin, E. M. Chudnovsky and T. Proctor, Phys. Rev. B
{\bf 88}, 224418 (2013).
\bibitem{PGC14}
T. C. Proctor, D. A. Garanin and E. M. Chudnovsky, Phys. Rev. Lett.
{\bf 112}, 097201 (2014).
\bibitem{HB89}
K. Hui and A. N. Berker, Phys. Rev. Lett. {\bf 62}, 2507 (1989).
\bibitem{Har74}
A. B. Harris, J. Phys. C {\bf 7}, 1671 (1974).


\end{thebibliography}


\end{document}